\documentclass[prc,twocolumn,showpacs,floatfix,nofootinbib,preprintnumbers,superscriptaddress,amsmath,amssymb]{revtex4-1}

\usepackage{CJKutf8}
\usepackage{epsfig}
\usepackage{color}
\usepackage{rotating}
\usepackage{hyperref}
\usepackage{natbib}
\usepackage{verbatim}
\usepackage{epstopdf}
\usepackage{graphicx}

\begin{document}

\begin{CJK*}{UTF8}{gbsn}

\title{Theoretical study of triaxial shapes of  neutron-rich Mo and Ru nuclei}

\author{C. L. Zhang (张春莉)}
\affiliation{%
Department of Physics and Astronomy and NSCL/FRIB Laboratory, Michigan State University, East Lansing, Michigan 48824, USA
}
\affiliation{%
State Key Laboratory of Nuclear Physics and Technology, School of Physics, Peking University, Beijing 100871, China
}

\author{G. H. Bhat}
\affiliation{%
Department of Physics, University of Kashmir, Srinagar, 190 006, India
}

\author{W. Nazarewicz}
\affiliation{%
Department of Physics and Astronomy and NSCL/FRIB Laboratory, Michigan State University, East Lansing, Michigan 48824, USA
}
\affiliation{%
Physics Division, Oak Ridge National Laboratory, Oak Ridge, Tennessee 37831-6373, USA
}
\affiliation{Institute of Theoretical Physics, Faculty of Physics, University of Warsaw, Pasteura 5, PL-02-093 Warsaw, Poland}

\author{ J. A. Sheikh}
\affiliation{%
Department of Physics, University of Kashmir, Srinagar, 190 006, India
}

\author{Yue Shi (石跃)}
\affiliation{%
Department of Physics and Astronomy and NSCL/FRIB Laboratory, Michigan State University, East Lansing, Michigan 48824, USA
}

\date{\today}

\begin{abstract}
\begin{description}
\item[Background] Whether atomic nuclei can possess triaxial shapes at their ground states is still a subject of ongoing debate. According to theory, good prospects for low-spin triaxiality are in the neutron-rich Mo-Ru region. Recently, transition quadrupole moments in rotational bands of even-mass 
neutron-rich isotopes of molybdenum and ruthenium nuclei have been measured. The new data have provided a challenge for theoretical descriptions invoking stable triaxial deformations.

\item[Purpose] To understand experimental data on rotational bands in the neutron-rich
Mo-Ru region, we carried out theoretical analysis of moments of inertia, shapes, and transition quadrupole moments of neutron-rich even-even  nuclei around $^{110}$Ru
using self-consistent mean-field and shell model techniques.

\item[Methods] To describe yrast structures in Mo and Ru isotopes, we use nuclear Density Functional Theory (DFT) with the optimized energy density functional UNEDF0. We also apply
Triaxial Projected Shell Model (TPSM)  to describe yrast and positive-parity, near-yrast band structures.

\item[Results] Our self-consistent DFT calculations predict triaxial ground-state deformations in  $^{106,108}$Mo and $^{108.110,112}$Ru and reproduce the observed low-frequency behavior of moments of inertia.  
As the rotational frequency increases, a negative-$\gamma$ structure,
associated with  the aligned $\nu(h_{11/2})^2$ pair, becomes energetically  favored.  The
computed transition quadrupole moments vary with angular momentum,  which reflects  deformation
changes with rotation; those variations are consistent with experiment.  
The  TPSM calculations explain the observed  band structures  assuming stable triaxial shapes.

\item[Conclusions] The structure of neutron-rich even-even  nuclei around $^{110}$Ru  is consistent with triaxial shape deformations. Our DFT and TPSM frameworks provide a consistent and complementary description of experimental data.
\end{description}

\end{abstract}


\maketitle

\end{CJK*}

\section{Introduction}

Neutron-rich molybdenum and ruthenium isotopes are known to exhibit shape changes and  shape-coexistence phenomena \cite{Zrregion,Stachel82,Srebrny06,Luo09,Lipska12,Smith12}.
With increasing neutron number, triaxial deformation is expected to appear in their ground states due to the occupation of 1$\nu h_{11/2}$ and 1$\pi g_{9/2}$
intruder orbitals \citep{Skalski97}. 

Experimentally, the clearest signature of triaxial shapes comes from the
$\gamma$-ray spectroscopy of rotating nuclei \citep{Bohr75,Frauendorf01}. 
Following Ref.~\citep{Cheifetz70}, which reported  evidence for rotational-like behavior
in the very neutron-rich even-even Zr-Pd region, numerous experiments
were devoted to investigations of shape transitions and rotational properties
in this region. The first systematic high-spin study of collective band structures   was undertaken in Ref.~\citep{Hotchkis91}, which reported  deformed configurations  in  $^{103,104,107}$Zr and  $^{107,108}$Mo and $^{108}$Mo.

The early work \citep{Stachel82} on $^{104}$Ru was indicative of a  transition from spherical to  triaxial shapes.  The collective nature of neutron-rich nuclei and triaxiality in  $^{110}$Ru and $^{112}$Ru  was confirmed in Ref.~\citep{Aysto90}
by reporting  a steady decrease of the  $\gamma$-band bandhead energy. More evidence of  collective triaxial behavior of Ru isotopes came from the spectroscopy of fission fragments \citep{Shannon94,Lu95,HY04,Wu06,Hamilton09,Hamilton10}.
The decrease of transition quadrupole moments at high spin showed that
the triaxial deformation in neutron-rich Mo isotopes could be spin-dependent
\citep{Smith96}. Another piece of experimental information on  triaxial deformations came from  the measurement of  the quasi-$\gamma$ band in
$^{110}$Mo \citep{Urban04,Watanabe11}. A more detailed information on quadrupole collectivity was obtained by Coulomb excitation studies, which succeeded in determining  unique sets of E2 and M1 matrix elements in $^{104}$Ru~\citep{Srebrny06} and $^{110}$Mo~\citep{Srebrny06} and extracting triaxial deformations using the collective quadrupole invariant approach. Rotational bands in $^{106}$Mo and  $^{108,110,112}$Ru
were investigated in Ref.~\citep{Zhu07,Luo09}, which reported chiral doublets associated with triaxial nuclear rotation. In recent papers \citep{Smith12,Snyder13}, transition quadrupole moments of
rotational bands in neutron-rich, even mass $^{102-108}$Mo and $^{108-112}$Ru
nuclei were measured for the spin range of 8-16\,$\hbar$, suggesting 
$\gamma$-softness effects or even triaxiality in these nuclei.

Theoretically,  triaxial ground states  in this region have been investigated  with different models. 
In  Ref.~\citep{Skalski97}, based on
macroscopic-microscopic approach, triaxial ground-state (g.s.)  minima were found in 
the neutron-rich  Mo isotopes with N=62-66 and also  in Ru isotopes.
In a systematic survey of Refs.~\citep{Moller06,Moller08},  the largest shell effects due to triaxial deformations were found around  $^{108}$Ru. The interacting boson model analysis of Ref.~\cite{Boyuk10} did not find any candidates for stable triaxiality in this region.  In
Ref.~\citep{Nomura10}, potential energy surface (PES) calculations for Ru
isotopes were carried out with Hartree-Fock (HF)  and interacting boson
models, and shallow  triaxial minima were found 
for $N=64-70$ (see also Ref.~\cite{Bentley11}). In the self-consistent Hartree-Fock Bogoliubov  study of Ref.~\citep{Guzman10} with Gogny D1S interaction, triaxial deformations were predicted 
for even-even isotopes $^{104-110}$Mo and $^{104,106}$Ru.

Collective rotation has been shown  
to enhance triaxial minima  in even-even  Mo and Ru isotopes \citep{Skalski97,Xu02,Faisal10}. For that reason, those nuclei are candidates for the presence of novel collective modes associated with triaxial rotation,
such as  wobbling motion and chiral bands
\citep{Bohr75,Frauendorf01}. The angular momentum alignment pattern in the lowest bands of Mo and Ru isotopes  was also explored within the projected shell model approach \cite{Bhat12,Liu13}. 
The axial study  \cite{Bhat12} provided a reasonable description  of yrast spectra and electromagnetic properties of $^{100-112}$Ru. This work was extended in Ref.~\cite{Liu13}, which also contains TPSM analysis of $^{110}$Mo and $^{114}$Ru. In the case of $^{114}$Ru, stable $\gamma$ deformation turned out to be crucial for reproducing the data.

In an attempt to explain the recent data on transition quadrupole moments in Mo and Ru nuclei, cranked relativistic
mean-field calculations \citep{Snyder13} predicted  axial prolate and oblate
ground states in those nuclei. However, the angular momentum dependence of resulting  transition quadrupole moments was not consistent with observations. As concluded in Ref.~\cite{Snyder13}: 
{\it Attempts to describe the observations in
mean-field based models, specifically cranked relativistic Hartree-Bogoliubov
theory, illustrate the challenge theory faces and the difficulty to infer information
on $\gamma$-softness and triaxiality from the data.}
To shed some light on this puzzle, and to further 
explore the importance of triaxial deformation in this mass region, 
we  apply the cranked self-consistent Hartree-Fock-Bogoliubov (CHFB)
method and TPSM to
the rotational properties of neutron-rich, even-even Mo and Ru
isotopes.

This article is organized as follows. Section~\ref{model} gives a brief introduction
to CHFB  and TPSM models used in this work. In Sec.~\ref{results}, the results of calculations for Mo and Ru
isotopes are presented and compared with experiment. Therein, we discuss 
potential energy and routhian  surfaces,  quasi-particle routhian spectra, 
and equilibrium deformations. To test the stability of CHFB minima with respect to angular momentum orientation, we carry our  tilted-axis cranking calculations employing the Kerman-Onishi conditions.
Finally, the conclusions of this work are given in  Sec.~\ref{summary}.

\section{The model}
\label{model}

\subsection{Cranked Skyrme-Hartree-Fock-Bogoliubov Model}

Our  CHFB calculations were performed with the DFT
solver {\sc HFODD} (version 2.49t) \citep{Schunck12}. Parity and $y$-signature ($\hat{R}_y=\exp(-i\pi \hat{J}_y)$) symmetries  are  conserved; the corresponding quantum numbers are denoted as $\pi$ and $r$.  The
quasi-particle HFB wave functions were expanded in  800 spherical
harmonic oscillator basis states with the oscillator frequency of
$\hbar\omega$ =49.2~MeV/A$^{1/3}$. We have tested that such
a basis provides a very reasonable precision for the observables studied.

In the particle-hole channel, we employ the  global Skyrme energy density functional
UNEDF0 optimized in Ref.~\citep{Kortelainen10}. In the pairing channel, we take the zero-range
density-dependent pairing force \citep{Dobaczewski01} with the
Lipkin-Nogami correction for particle-number fluctuations. The original pairing strengths are taken as $(V_0^{\nu}, V_0^{\pi})
= (-170.374, -199.202)$ MeV fm$^3$, with a cutoff energy in the quasi-particle
spectrum of $E_{\rm cut}=60$ MeV. In the present calculation, the strengths of
the pairing force for neutrons and protons have been increased by 5\% to reproduce
 the kinematic moment of inertia of the g.s.  band  (g.b.) of $^{106}$Mo. As discussed below,  the
calculated potential energy surfaces are not sensitive to such a small variation of 
pairing strengths.

In the multi-dimensional potential energy surface calculations, the constraints
are imposed on expectation values of multipole moments. We use the Augmented
Lagrangian Method \citep{Staszczak10} to perform the constrained iterations.  The total routhians were computed within the principal-axis cranking approach \cite{Frauendorf01}.  However, to study the stability of the resulting triaxial minima with respect to the orientation of the angular momentum vector, we applied the  Kerman-Onishi conditions implemented as in  Refs.~\cite{Shi12,Shi13}. Since the Lipkin-Nogami method is not strictly variational, the g.b. minimum at nonzero angular momentum was obtained by minimizing  the constrained total routhian surface. This increases precision of calculations, especially when the minima are soft \cite{Sto03}.

\subsection{Triaxial Projected Shell Model}

Recently, multi-quasiparticle TPSM approach
has been developed and it has been shown to provide a consistent
description of yrast, $\gamma$ ($K=2$) and $\gamma\gamma$ ($K=4$) bands in transitional
nuclei \citep{JS99,YK00}. In this method, the three dimensional projection
technique is employed to project out the good angular-momentum states from product states built upon quasiparticle (q.p.) configurations of
triaxially deformed Nilsson+BCS model. The shell model Hamiltonian is then
diagonalized in this  angular-momentum projected basis. The TPSM
space includes multi-quasiparticle  states; hence, is capable of describing
near-yrast  band structures at high-spins \citep{JY01,YJ02,GH08,JG09,Liu13}.

The TPSM basis employed in this study  consists of 0-q.p. vacuum, two-proton, two-neutron, and the
four-q.p. configurations \citep{GH08}. The q.p. basis chosen is adequate
to describe high-spin states up to angular momentum $I\sim 20$. In the present analysis we
shall, therefore, restrict our discussion to this spin regime.  
As in the earlier TPSM calculations, we use the pairing plus
quadrupole-quadrupole Hamiltonian \citep{KY95,Liu13}:
\begin{equation}
\hat H = \hat H_0 - {1 \over 2} \chi \sum_\mu \hat Q^\dagger_\mu
\hat Q^{}_\mu - G_M \hat P^\dagger \hat P - G_Q \sum_\mu \hat
P^\dagger_\mu\hat P^{}_\mu,
\label{hamham}
\end{equation}
where $\hat H_0$ is the single-particle spherical Nilsson Hamiltonian,    $\chi$ is the  strength of the quadrupole-quadrupole force related in a self-consistent way to deformation of the q.p. basis, and $G_M$ and $G_Q$ are the strengths of the monopole and quadrupole pairing terms, respectively.
The configuration space employed corresponds to  three principal oscillator shells ${\cal N}_{\rm osc}$: $\nu[3,4,5]$ and $\pi[2,3,4]$. The pairing strengths have been parametrized as in Refs.~\citep{sjb12,Bhat12} in terms of two constants $G_1$ and $G_2$. In this work, 
we choose  $G_1=16.22$\,MeV and $G_2=22.68$\,MeV; with these pairing strengths we 
approximately reproduce the experimental odd-even mass differences in this region.
The quadrupole pairing strength $G_Q$ is
assumed to be proportional to $G_M$, and the proportionality
constant was set to 0.18 \citep{sjb12,Bhat12}. The single-particle basis is that of the deformed Nilsson Hamiltonian parametrized in terms of  axial ($\varepsilon$) and triaxial  ($\varepsilon'$) deformations related to the
standard Bohr triaxiality parameter $\gamma$  by $\gamma=\arctan(\varepsilon'/\varepsilon)$.

\section{Results and Discussions}
\label{results}

\subsection{CHFB Results}
\label{SHFB-res}

\subsubsection{Ground-state Potential Energy Surfaces}

\begin{figure}[!htb]
\includegraphics[width=0.9\columnwidth]{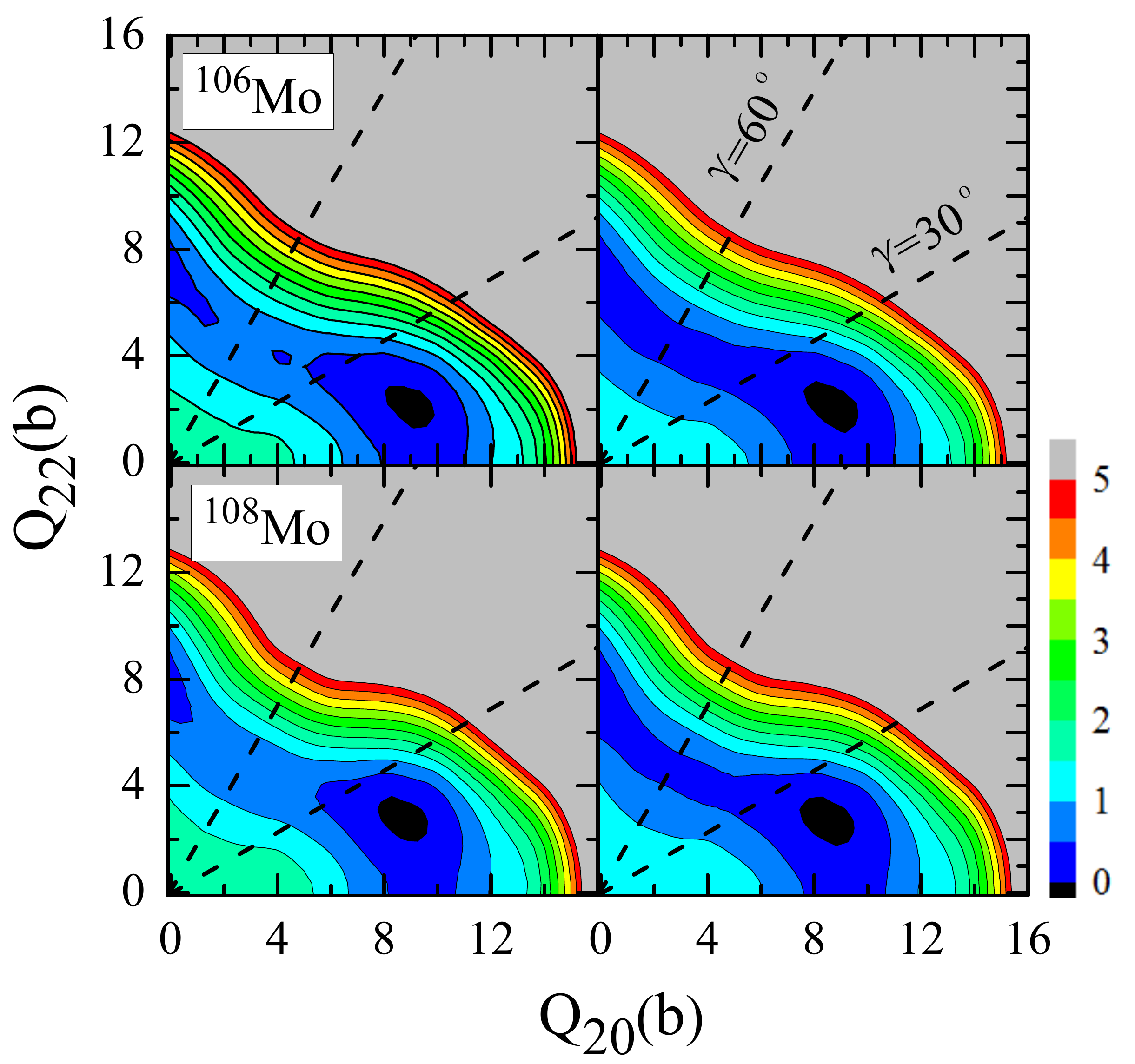}
\caption{(Color online) PES in $(Q_{20}, Q_{22})$ plane in CHFB+UNEDF0 for $^{106}$Mo and $^{108}$Mo. Left: standard pairing strengths. Right:  pairing strengths increased by 5\%, see text.
The difference between contour lines is  0.5 MeV.}
\label{mo-pes}
\end{figure}
\begin{figure}[!htb]
\includegraphics[width=0.9\columnwidth]{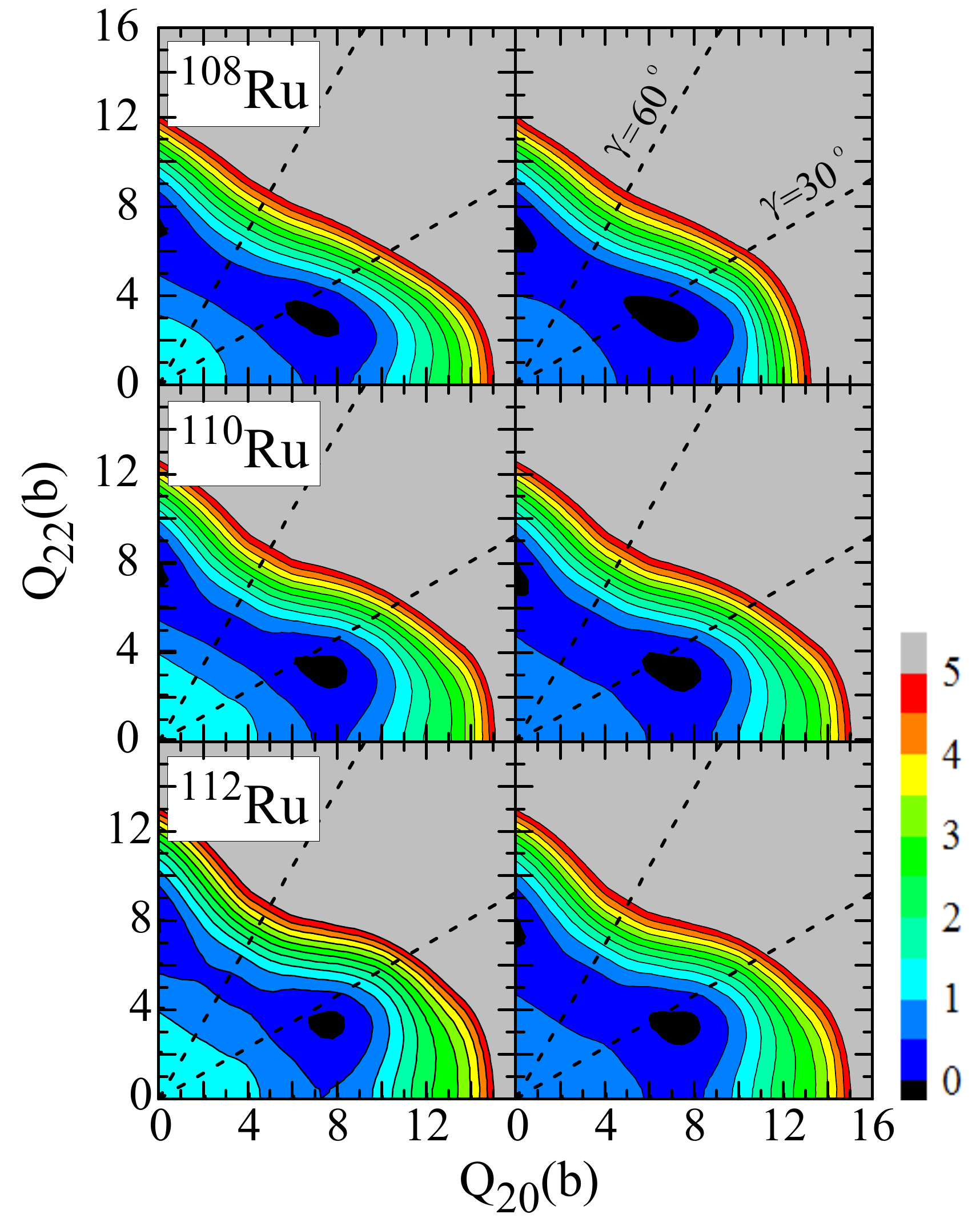}
\caption{(Color online) Similar to Fig.~\ref{mo-pes}, but for $^{108}$Ru, $^{110}$Ru, and $^{112}$Ru.}
\label{ru-pes}
\end{figure}

The g.s. UNEDF0 PESs  for $^{106,108}$Mo and $^{108,110,112}$Ru are shown in Figs.~\ref{mo-pes} and \ref{ru-pes}, respectively. All these nuclei are predicted to be triaxial in their ground states. It is seen that the PESs are practically not affected by a 5\% change in the pairing strengths. In particular, the triaxial
minima appearing at $(Q_{20}, Q_{22}) \approx (8.0-9.5, 2.0-3.0)$~b are not affected by pairing. The corresponding g.s. quadrupole deformations $(\beta_2, \gamma)$ are displayed in Table~\ref{tab0}.
\begin{table}[htb]
\caption{Bohr quadrupole deformation parameters
$\beta$ and $\gamma$ calculated in CHFB+UNEDF0  for $^{106,108}$Mo and $^{108,110,112}$Ru.}
\begin{ruledtabular}
\begin{tabular}{cccccc}  & $^{106}$Mo  &$^{108}$Mo &$^{108}$Ru & $^{110}$Ru & $^{112}$Ru \\
\hline $\beta_2$ & 0.19    &  0.18   & 0.16    & 0.16     & 0.15  \\
       $\gamma$ &  16$^\circ$     &  18$^\circ$     & 24$^\circ$      & 25$^\circ$        & 24$^\circ$  
\end{tabular}\label{tab0}
\end{ruledtabular}
\end{table}
For $^{106,108}$Mo, we predict the triaxial g.s. minima at
$(\beta_2, \gamma) \approx (0.19, 17^\circ)$. Similar results were obtained in the macroscopic-microscopic  calculations of Refs.~\citep{Galeriu86,Skalski97,Moller08} and HFB+D1S calculations~\citep{Guzman10}.
For $^{108,110,112}$Ru, we also predict  triaxial g.s. minima; this is consistent with  Refs.~\citep{Skalski97,Faisal10} and  HF+SIII calculations of Ref.~\citep{Aysto90}. Triaxial g.s. shapes for $^{108,110}$Ru were also obtained in the survey \cite{Moller08} but  $^{112}$Ru was calculated to be axial.

\subsubsection{High spin behavior}

\begin{figure}[!htb]
\includegraphics[width=0.9\columnwidth]{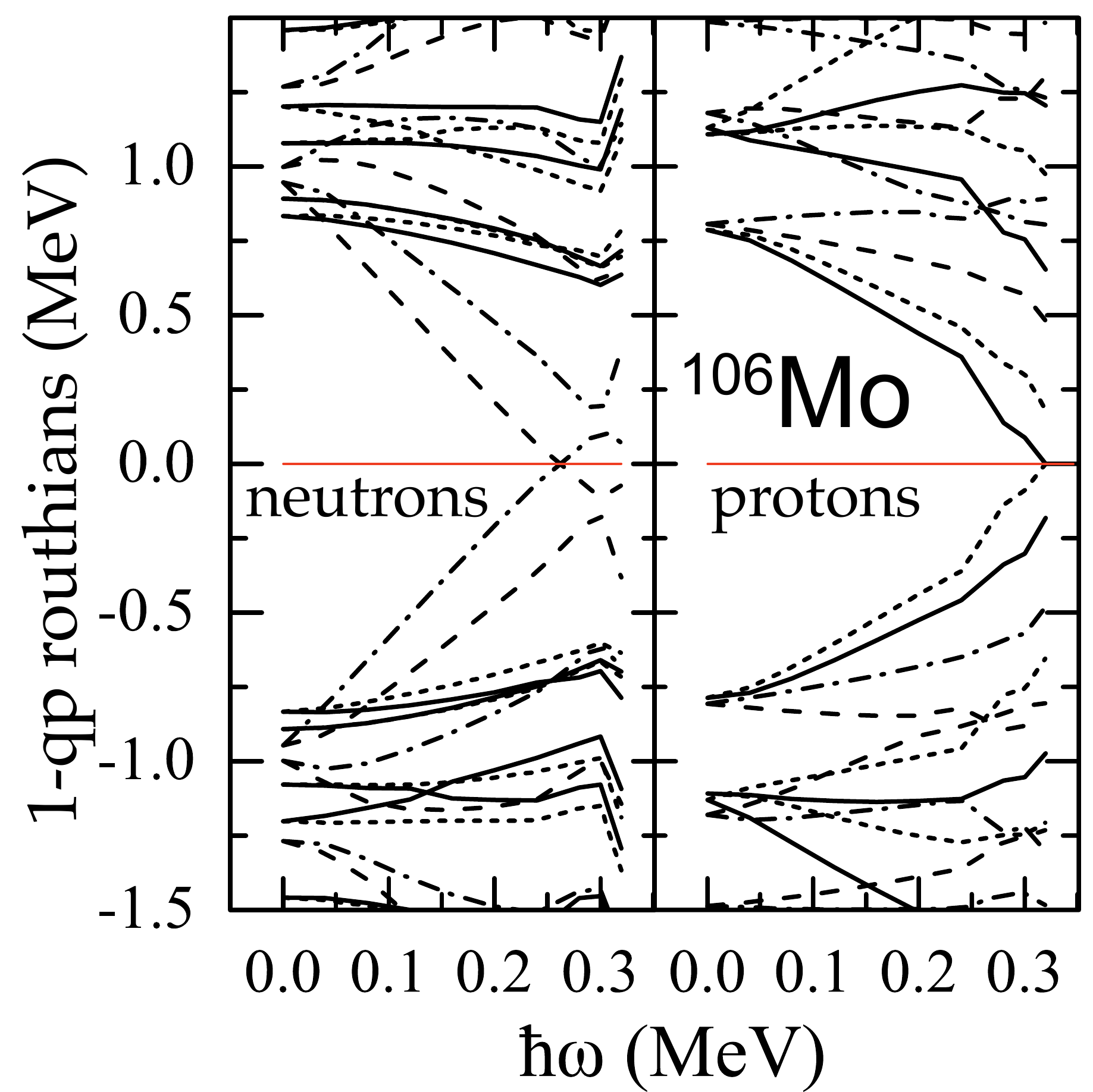}
\caption{(Color online) One-quasiparticle routhian diagram 
for $^{106}$Mo obtained with CHFB+UNEDF0.
The parity $\pi$ and
signature $r$ of individual levels are indicated in the following way:  $\pi=+,r=+i$ -- solid line;
$\pi=+,r=-i$ -- dotted line; $\pi=-,r=+i$ -- dot-dashed line; $\pi=-,r=-i$ -- dashed line. The thin line indicates the Fermi energy.
}
\label{mo106-quasi}
\end{figure}

\begin{figure}[!htb]
\includegraphics[width=0.9\columnwidth]{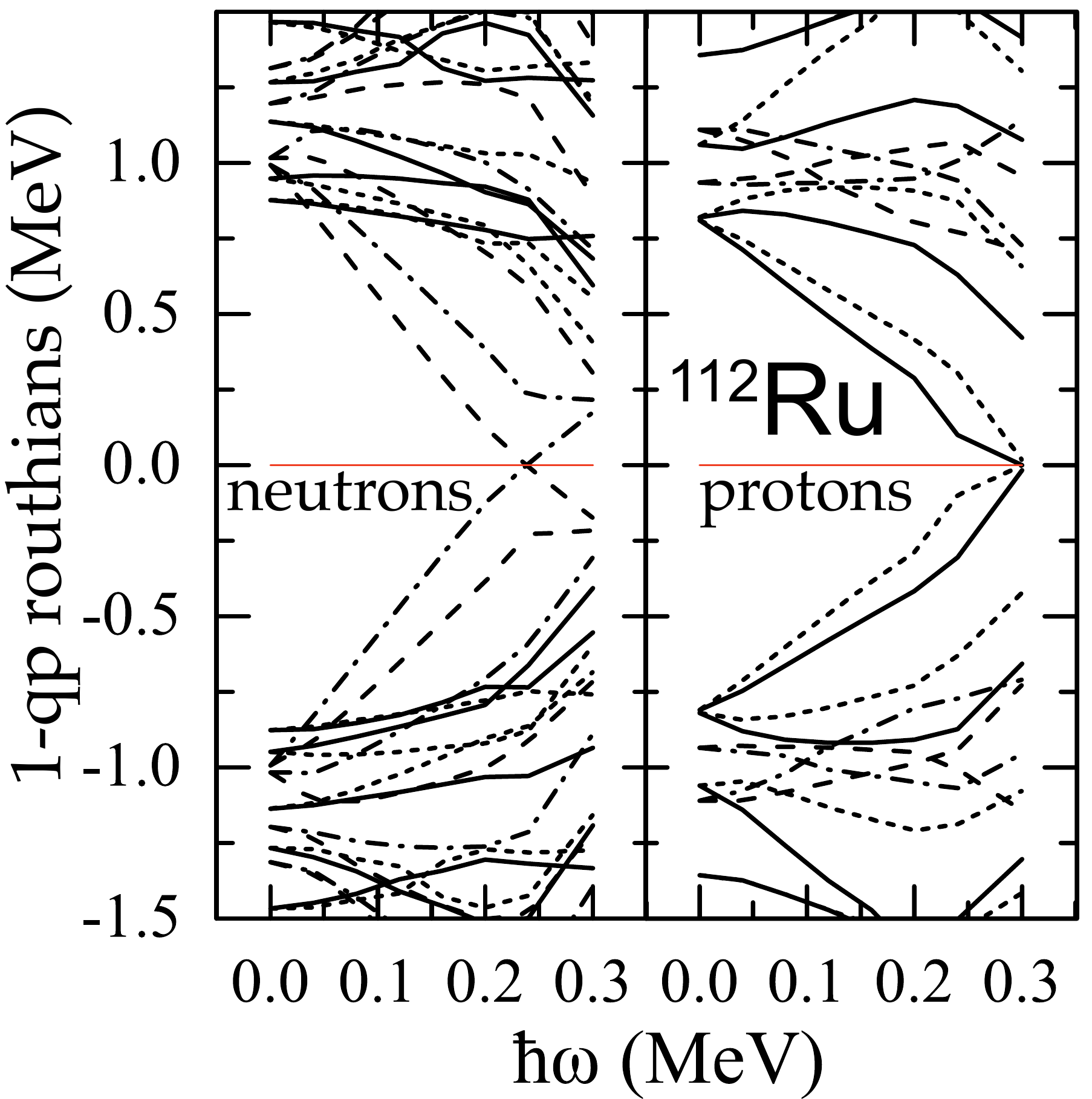}
\caption{(Color online) Similar to Fig.~\ref{mo106-quasi}, but for $^{112}$Ru.}
\label{ru112-quasi}
\end{figure}

The angular momentum alignment pattern  of Mo and Ru nuclei is governed by  the $\nu h_{11/2}$
and $\pi g_{9/2}$  high-$j$ q.p. excitations, which  give rise to strong shape polarization effects \citep{Skalski97}. 
Figures~\ref{mo106-quasi} and \ref{ru112-quasi} show self-consistent CHFB+UNEDF0
1-q.p. routhian diagrams versus rotational frequency  for $^{106}$Mo and $^{112}$Ru,
respectively. In both cases, the alignment of $\nu (h_{11/2})^2$ and $\pi (g_{9/2})^2$ pairs occurs at similar rotational frequencies of $\hbar \omega \approx$0.3\,MeV. At higher rotational frequencies, a transition is expected from the g.b. configuration to aligned
$\nu (h_{11/2})^2$ and $\pi (g_{9/2})^2$ 2-q.p. bands, and then to a 4-q.p.
$\nu (h_{11/2})^2\pi (g_{9/2})^2$ band. These two consecutive crossings are difficult to follow in CHFB calculations, as this would require a diabatic-configuration extension \cite{Cwiok80,Szymanski83,Axelsson02} of the current framework. Such an  extension is highly nontrivial in CHFB as the  self-consistent  mean-fields associated with aligned configurations are expected to differ significantly from those of the g.b. \citep{Skalski97}. Moreover, pairing correlations in the aligned bands are quenched and this causes numerical instabilities around the band crossing. Therefore, to provide interpretation of the transition quadrupole moments at higher angular momenta, we carry out cranked Skyrme-Hartree-Fock (CHF)
calculations without pairing at $\hbar \omega >0.3$ MeV. In this case, diabatic configurations 
can be defined by the number of single-particle routhians  occupied
in the four parity-signature blocks \citep{Dobaczewski00}. Specifically, each  neutron and proton configurations is defined by four occupation numbers $[N_{++},N_{+-},N_{-+},N_{--}]$ representing
the number of particles $N_{\pi,r}$ occupying single-particle states of given $\pi$ and $r$. The lowest total routhian with $\pi=+$ and $r=1$ is associated with the yrast configuration.

\begin{figure}[!htb]
\includegraphics[width=0.9\columnwidth]{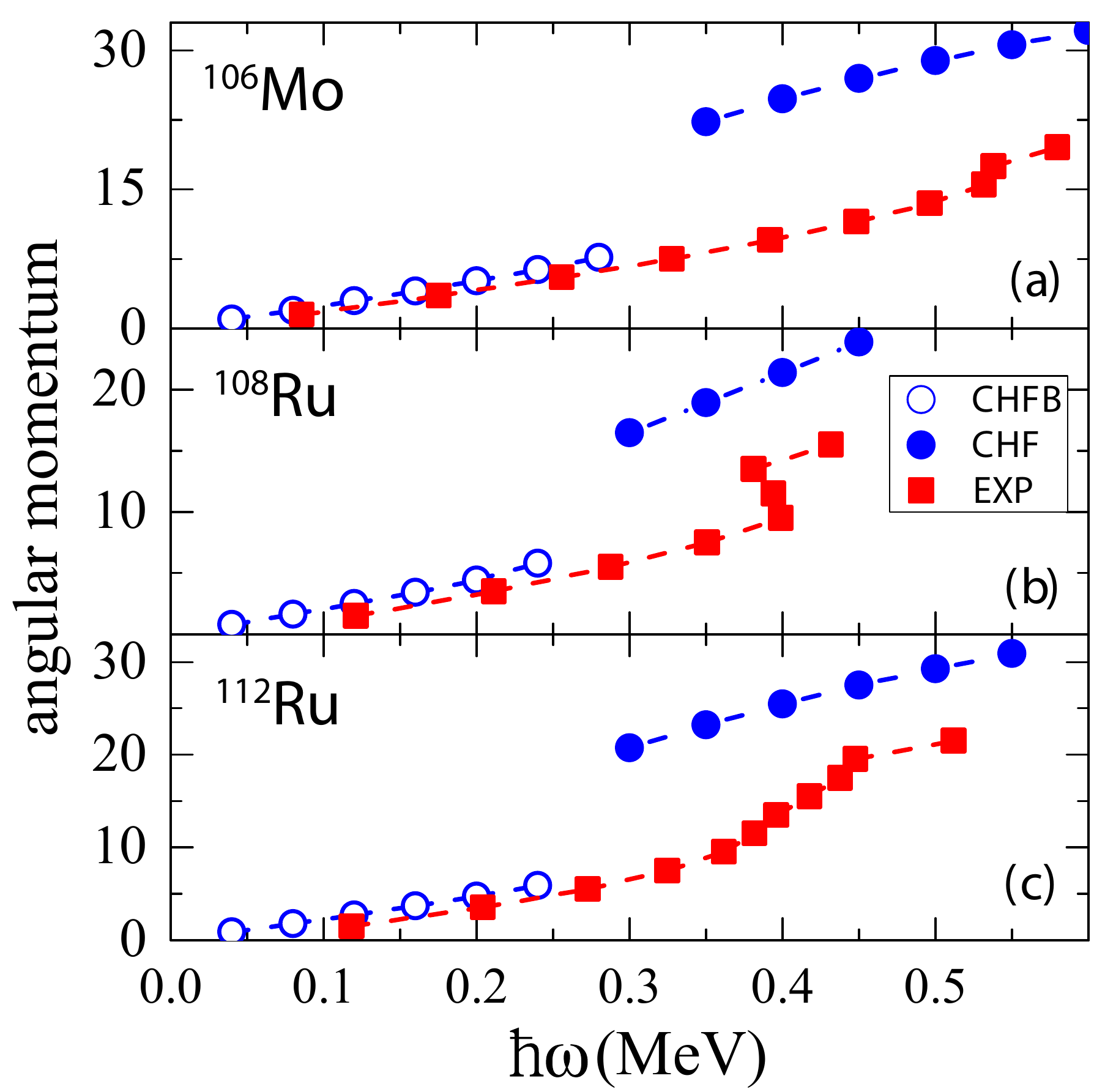}
\caption{(Color online) Angular momentum alignment for $^{106}$Mo and $^{108,112}$Ru. CHFB ($\hbar \omega <0.3$ MeV) and CHF ($\hbar \omega >0.3$ MeV) calculations are compared to experiment \cite{HY04,Lu95,Wu06}}
\label{spin-omega}
\end{figure}

The angular momentum alignment (total angular momentum as a function of
rotational frequency) is shown in  Fig.~\ref{spin-omega} for $^{106}$Mo, $^{108}$Ru, and $^{112}$Ru. Below the predicted band crossing at  $\hbar\omega\approx$0.3 MeV, our calculations reproduce 
experiment well. (Note that our pairing strengths were
adjusted to match the kinematic moment of inertia of $^{106}$Mo.) The first band crossing, associated with the alignment of the $h_{11/2}$ neutron pair, is seen in $^{108,112}$Ru data slightly below $\hbar\omega=0.4$\,MeV, and it is significantly delayed in $^{106}$Mo.  The predicted aligned configuration above the band crossing  has a fairly different shape as compared to that of the g.b., and it is difficult to follow the g.b. at $\hbar \omega >0.3$\,MeV.

\begin{figure}[!htb]
\includegraphics[width=0.9\columnwidth]{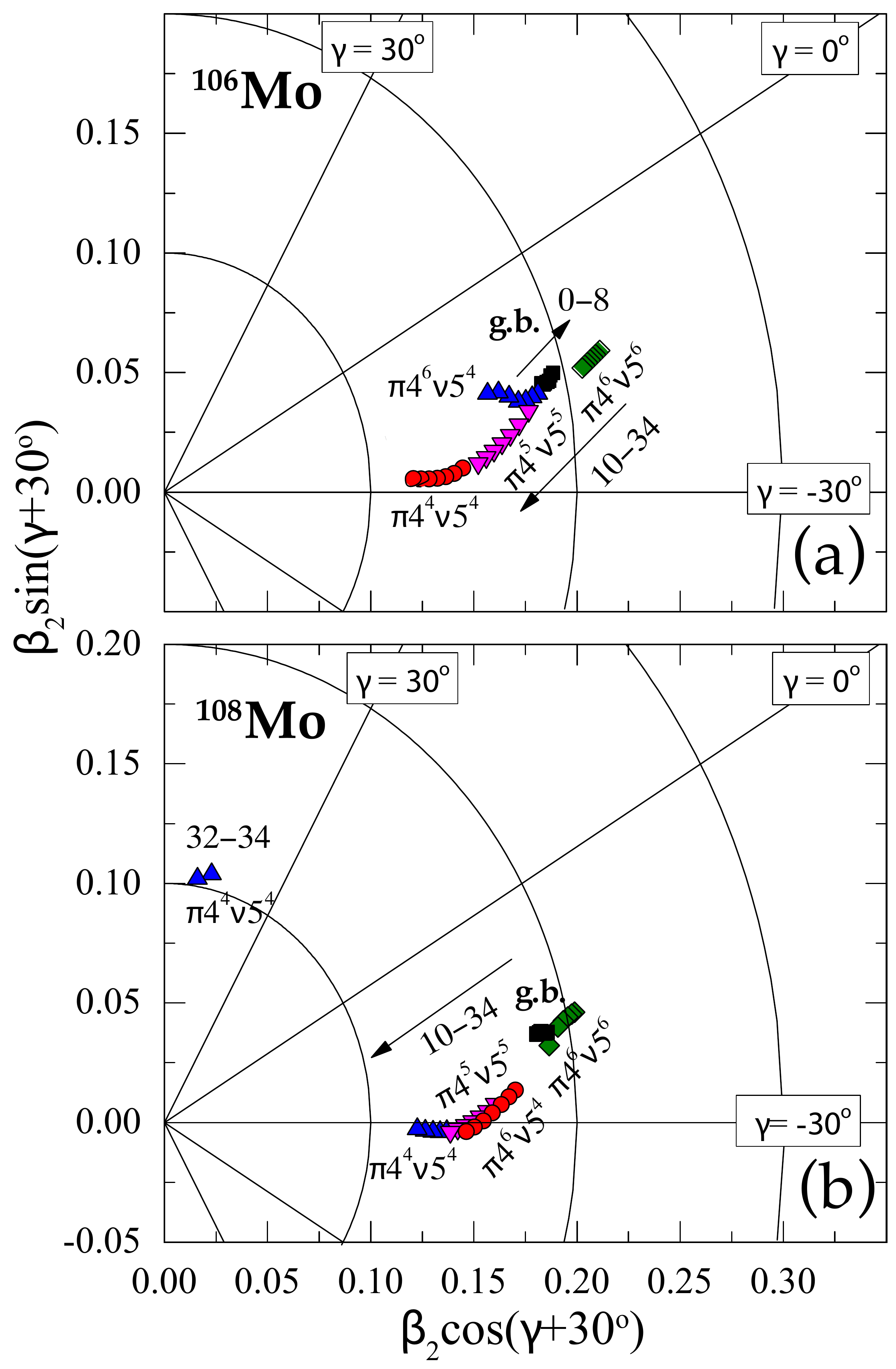}
\caption{(Color online) Summary of equilibrium deformations of the lowest $\pi=+, r=1$
bands in $^{106,108}$Mo calculated with CHFB+UNEDF0 (ground band) and CHF+UNEDF0 (aligned bands).  The rotational frequency is varied from zero to $\hbar\omega = 0.6$\,MeV. The corresponding range of angular momentum is indicated. The aligned band are classified according to the number of occupied high-${\cal N}$ intruder levels (${\cal N}= 5$ and 4 for neutrons and protons, respectively). 
}
\label{mo-beta-gamma}
\end{figure}

\begin{figure}[!htb]
\includegraphics[width=0.9\columnwidth]{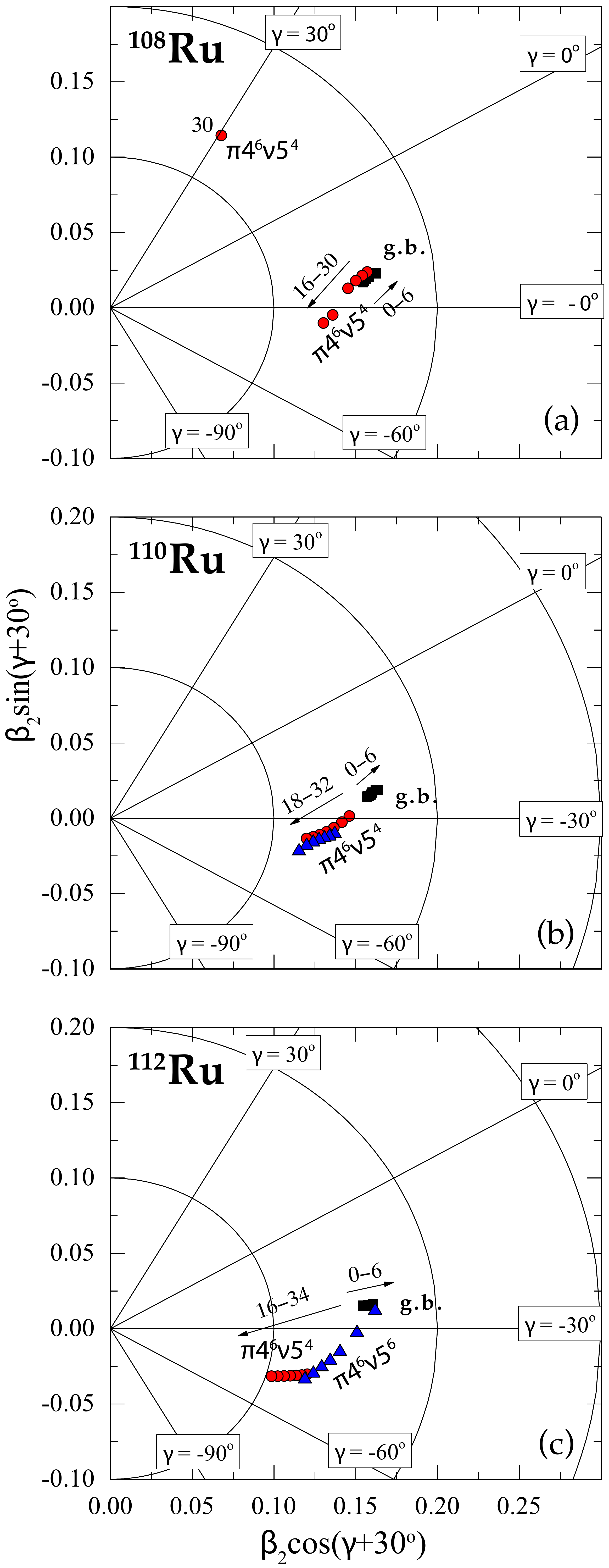}
\caption{(Color online) Similar to Fig.~\ref{mo-beta-gamma}, but for
 $^{108,110,112}$Ru. 
}
\label{ru-beta-gamma}
\end{figure}

To investigate  the evolution of nuclear shapes with rotation, we compute the
equilibrium $\beta_{2}$ and $\gamma$ deformations for low-lying $\pi=+, r=1$
bands in  $^{106,108}$Mo (Fig.~\ref{mo-beta-gamma}) and
$^{108,110,112}$Ru (Fig.~\ref{ru-beta-gamma}). In all cases considered, the triaxial paired g.b. undergoes small centrifugal stretching in the direction of $\beta_2$. For instance, in the case of  $^{108}$Ru,  $\beta_2$ increases from the value of 0.15 at $\hbar\omega=0$ to 0.17 at $\hbar\omega=0.3$\,MeV. 

At higher spins ($10 \le I \le 36$), when  pairing is neglected in our calculations,  it is useful
to label many-body configurations $[N_{++},N_{+-},N_{-+},N_{--}]$ by
the number of occupied intruder levels, i.e., ${\cal N}_{\rm osc}= 4$ protons
(primarily $g_{9/2}$) and ${\cal N}_{\rm osc}= 5$ neutrons
(primarily $h_{11/2}$). For instance, the aligned configuration 
$\pi(9,9,12,12)\otimes\nu(17,17,15,15)$ in $^{106}$Mo (shown by circles in Fig.~\ref{mo-beta-gamma}(a)) can be denoted as $\pi 4^4 \nu 5^4$, and the same
label applies to the 
$\pi(9,9,12,12)\otimes\nu(18,18,15,15)$ in $^{108}$Mo
(shown by up-triangles in Fig.~\ref{mo-beta-gamma}(b)). 

The quadrupole deformations $\beta_2$ of aligned bands  
are predicted to be in the range of $0.12 \le \beta_2 \le 0.16$, which represents a  reduction as compared to the shapes of paired ground-state bands. The aligned bands 
remain triaxial with $\gamma$ values  around $-30^{\circ}$ up to $\hbar\omega=0.6$\,MeV. This finding is consistent with the results of early Ref.~\cite{Skalski97} employing cranked macroscopic-microscopic approach.
At  the highest rotational frequencies considered, our calculations predict the appearance of aligned triaxial configurations with $\gamma>0$, which eventually terminate at oblate shapes ($\gamma \approx 60^{\circ}$), see, e.g., Fig.~\ref{mo-beta-gamma}(b).

\begin{figure}[!htb]
\includegraphics[width=0.9\columnwidth]{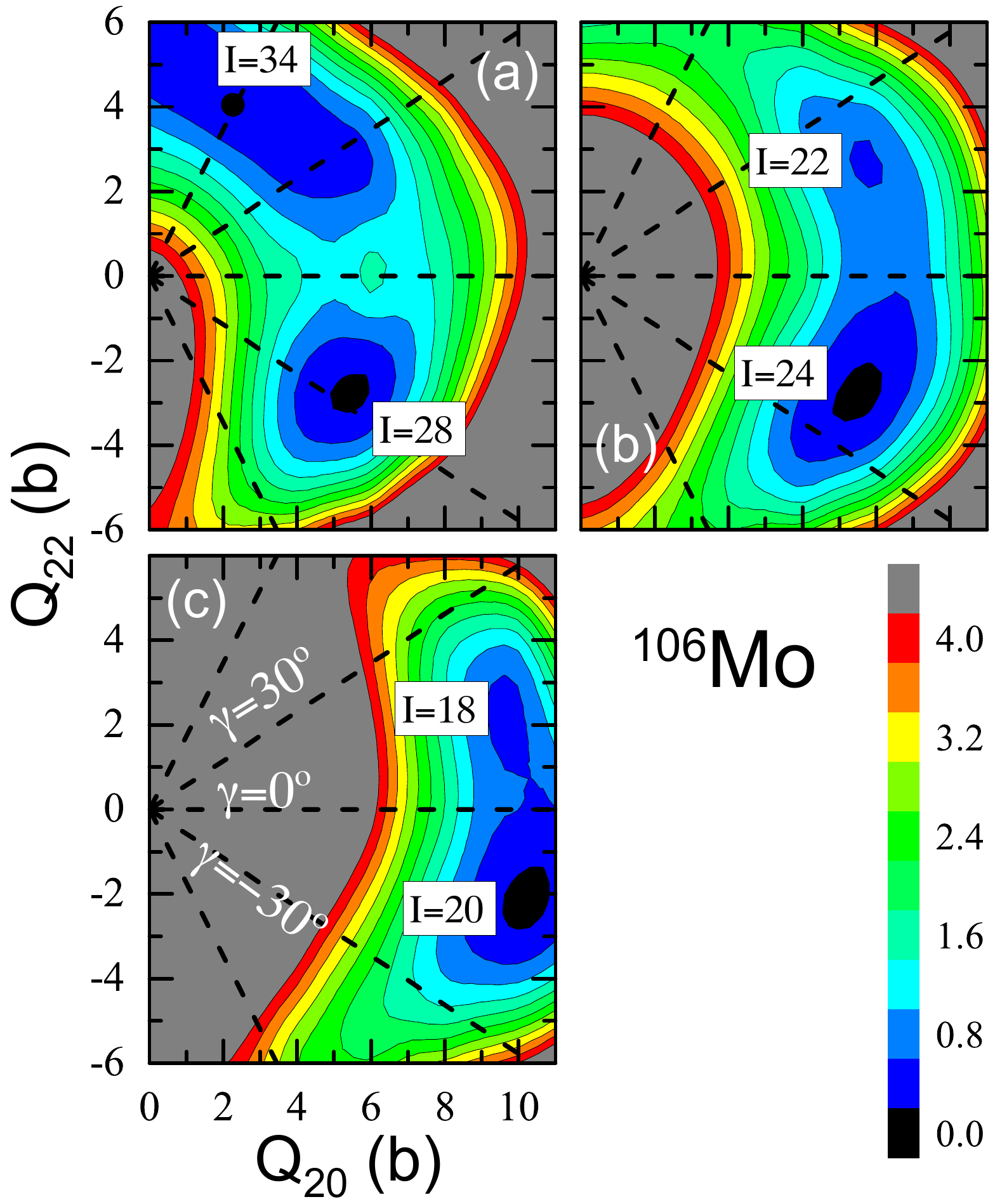}
\caption{(Color online) Diabatic total routhian surfaces for $^{106}$Mo at $\hbar\omega = 0.5$ MeV calculated in CHF+UNEDF0 for the configurations:
(a) $\pi(9,9,12,12)\otimes \nu(17,17,15,15)$ ($\pi 4^4 \nu 5^4$ in Fig.~\ref{mo-beta-gamma}(a));
(b) $\pi(10,10,11,11)\otimes \nu(17,17,15,15)$ ($\pi 4^6 \nu 5^4$); and 
(c) $\pi(10,10,11,11)\otimes \nu(16,16,16,16)$ ($\pi 4^6 \nu 5^6$).
}
\label{mo106-trs}
\end{figure}

To study the stability of different triaxial minima at high spins, we analysed related diabatic total routhians in the ($Q_{20}, Q_{22}$) plane. 
In  Fig.~\ref{mo106-trs} we show the total routhian surfaces at $\hbar\omega=0.5$ MeV for the selected low-lying  aligned configurations
in $^{106}$Mo discussed in Fig.~\ref{mo-beta-gamma}(a). For all those configurations, the collective  triaxial minimum with $\gamma$  between  $-30^\circ$ 
and $-15^\circ$ appears  lowest in energy. For the configuration $\pi 4^4 \nu 5^4$ shown in Fig.~\ref{mo106-trs}(a), we also predict a noncollective oblate state with $I=34$ that represents a  termination point of $\gamma>0$ band.

To eliminate spurious minima that are unstable with respect to the angular momentum orientation, we also investigated the dependence of the routhians  on the angular momentum tilting angle  $\theta$ with respect to the axis of rotation ($y$-axis). To this end, we used  the tilted-axis-cranking formalism of Refs.~\cite{Shi12,Shi13}.
The  calculations were performed for the aligned bands in $^{106}$Mo   At $\hbar\omega<0.5$\,MeV, the total routhians
of triaxial ($\gamma<0$) configurations  $\pi 4^4 \nu 5^4$, $\pi 4^5 \nu 5^5$,  and $\pi 4^6 \nu 5^4$  of  Fig.~\ref{mo106-trs}(a)
show a minimum at $\theta=0^\circ$. At $\hbar\omega\approx 0.5$\,MeV, the routhians become very soft in $\theta$, indicating a large-amplitude collective motion in this direction. This instability is not present for a $(\pi=-, r=1)$ configuration
$\pi(9,9,12,12)\otimes \nu(18,17,15,14)$ ($\pi 4^4 \nu 5^5$), which shows a pronounced minimum 
at  $\theta=90^\circ$ associated with $\gamma > 0$. This result is consistent with the deformation-driving effect of aligned $h_{11/2}$ neutrons orbitals discussed in Ref.~\cite{Skalski97}.

\begin{figure}[!htb]
\includegraphics[width=0.9\columnwidth]{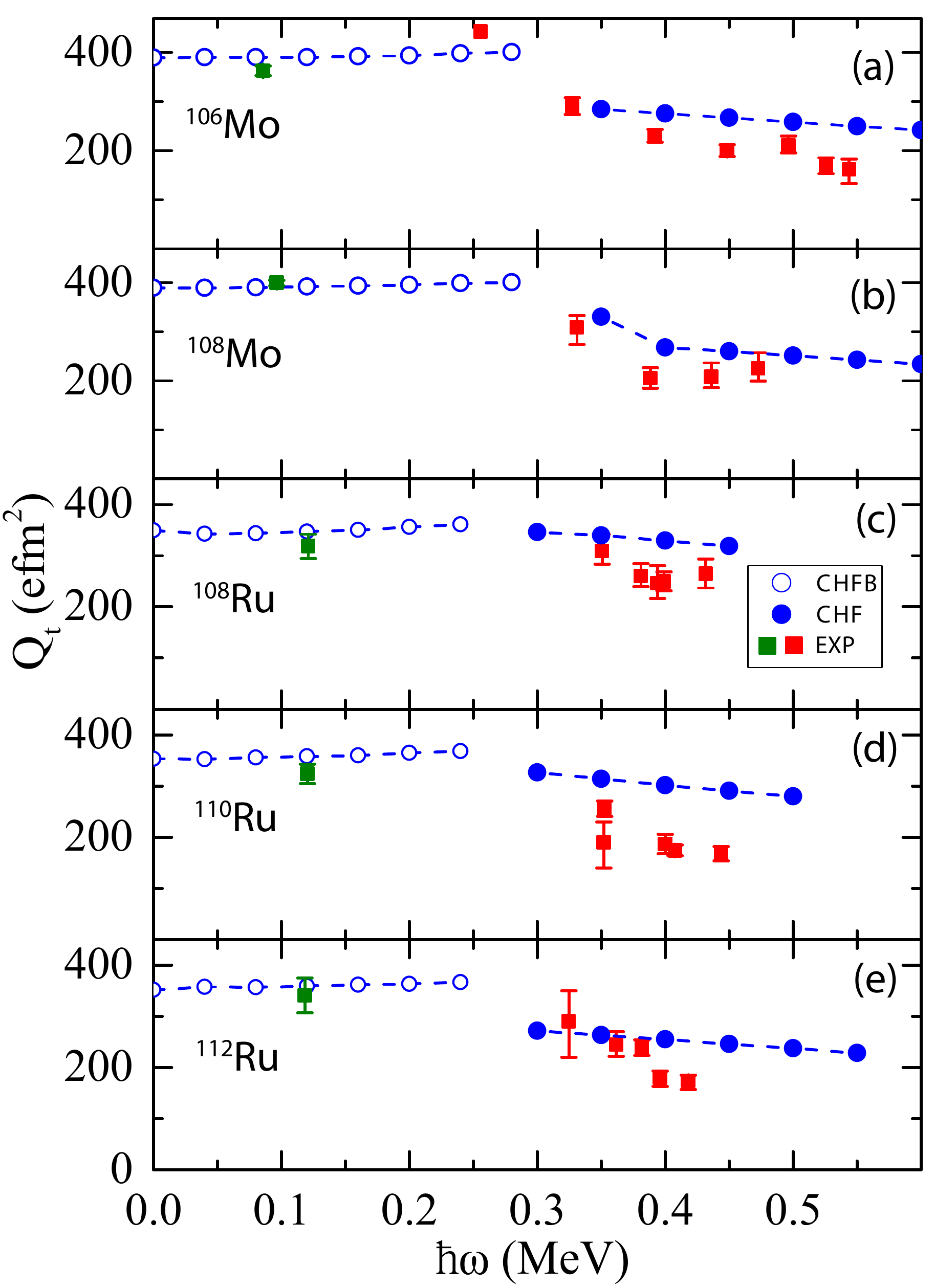}
\caption{(Color online) Transition quadrupole moments for $^{106,108}$Mo and  $^{ 108,110,112}$Ru
calculated in CHFB  (open circles) and CHF (dots) compared to experiment. The  $Q_t$ value  at $I=2$ is taken from  Ref.~\citep{Raman87} and  the high-spin values  from Ref.~\citep{Snyder13}.
}
\label{qt-omega}
\end{figure}

The transition quadrupole moments  along the yrast band in $^{106,108}$Mo and $^{108,110,112}$Ru are shown in  Fig.~\ref{qt-omega} as a function of
rotational frequency. At low rotational frequencies $\hbar\omega<0.3$\,MeV, there is a gradual increase of  $Q_{t}$ with $\omega$ due to the centrifugal stretching effect seen  in Figs.~\ref{mo-beta-gamma} and \ref{ru-beta-gamma}.
As discussed earlier, at higher frequencies cranking calculations are performed
without pairing. While this approximation seriously affects the predicted angular momentum alignment shown in Fig.~\ref{spin-omega}, the equilibrium shapes obtained in the CHF are reasonable approximations to those obtained in the full CHFB framework \cite{Naz85,Dud87}, and reproduce experimental $Q_t$-values for aligned configurations \cite{Sat96,Mat07}.
As seen in Fig.~\ref{qt-omega}, the predicted transition
quadrupole moments in aligned bands are slightly reduced with respect to the low-spin region due to
the deformation reduction associated with the  aligned $\nu h_{11/2}$ and $\pi g_{9/2}$ pairs.
This reduction is generally consistent with experiment, except perhaps for  $^{110}$Ru, where theory overestimates the measured $Q_t$ values above $\hbar\omega=0.3$\,MeV.

\subsection{TPSM Results}
\label{TPSM-res}

The TPSM calculations proceed in several stages. In the first stage,
the deformed basis is constructed based on the eigenstates  of the
triaxially deformed Nilsson potential. The $\beta_2$ deformation
has been chosen such that the lowest quadrupole transition from $2^+ \rightarrow 0^+$ is
reproduced. The non-axial deformation parameter $\gamma$ is chosen
from the minimum of the  g.s.  PES obtained in TPSM calculations. For
 $^{108}$Mo and $^{108,110}$Ru, where  the PES minima are $\gamma$-soft, triaxial deformation was adjusted to  the
experimental bandhead energy of the $\gamma$-band as it is known to be
quite sensitive to the $\gamma$-deformation. The adopted values of $\gamma$ are listed in Table~\ref{tab1}. It is seen that strongly triaxial shapes are expected in all cases, and this confirms the CHFB+UNEDF0 results.
%
\begin{table}
\caption{Triaxial quadrupole deformation parameters
 $\gamma$ employed in the TPSM calculation
for $^{106,108}$Mo and $^{108,110,112}$Ru isotopes.}
\begin{ruledtabular}
\begin{tabular}{cccccc }
 Nucleus   &    $^{106}$Mo  &$^{108}$Mo &$^{108}$Ru & $^{110}$Ru & $^{112}$Ru \\\hline  
  $\gamma$  & 20$^\circ$      &  25$^\circ$     & 29$^\circ$      & 28$^\circ$        & 25$^\circ$ 
\end{tabular}\label{tab1}
\end{ruledtabular}
\end{table}

In the next step, the good angular-momentum basis is obtained from the
triaxial Nilsson+BCS  wave functions by  using the three-dimensional
angular-momentum projection operator. Finally,  the shell model
Hamiltonian (\ref{hamham}) is diagonalized in this good-angular-momentum basis.

\subsubsection{Band structures}

\begin{figure}[!htb]
\includegraphics[width=0.9\columnwidth]{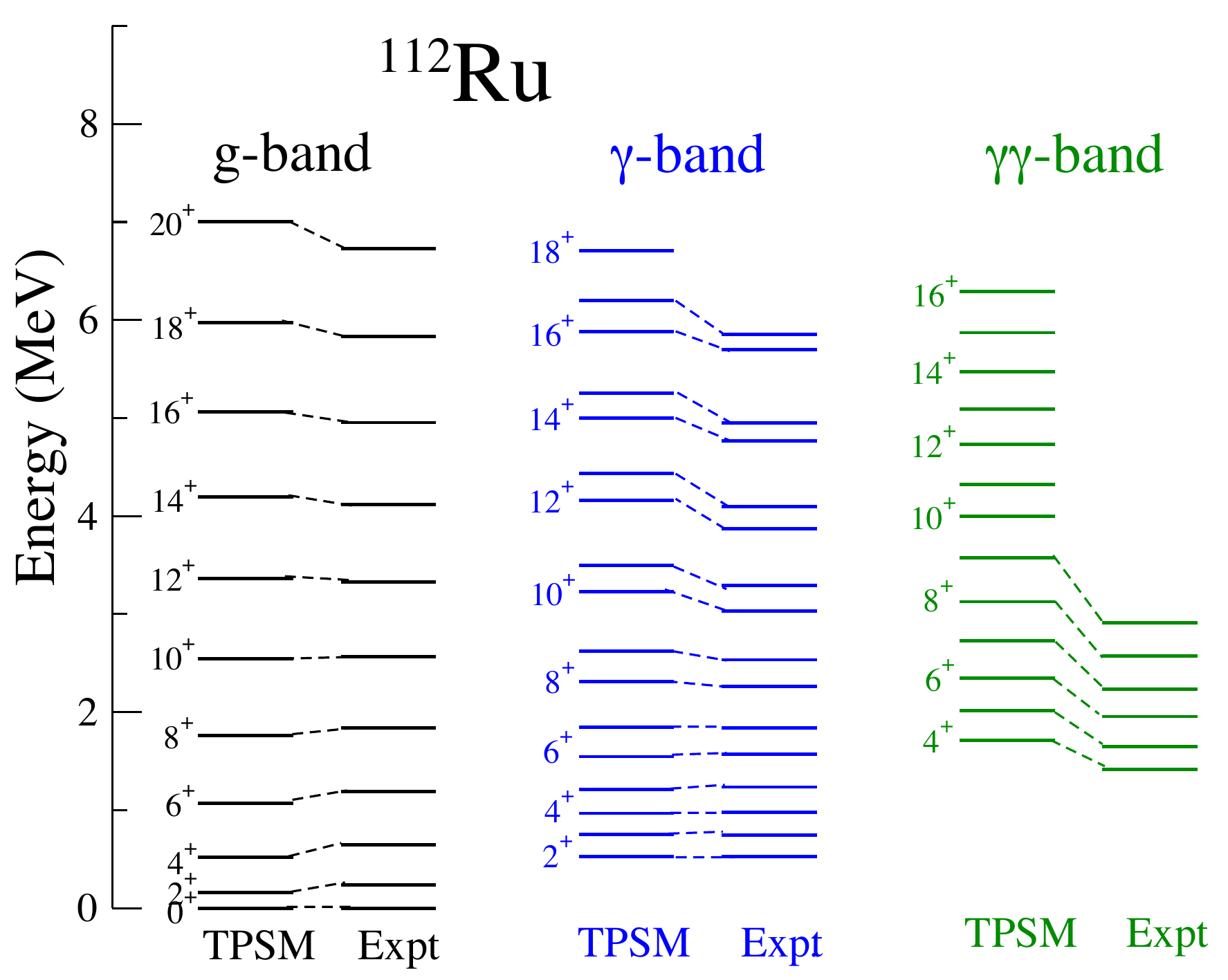}
\caption{(Color online) Comparison between experimental \citep{112ru} and calculated band
structure of $^{112}$Ru.}
\label{fig11}
\end{figure}
The TPSM band structures and the associated electromagnetic transition rates obtained in TPSM are quite rich and will be discussed in a separate paper \cite{BhatTBP}. Figure Fig.~\ref{fig11} shows the results for $^{112}$Ru as an representative example, as in this nucleus all bands  are known up to high spins. It is seen that TPSM
reproduces the  experimental band energies quite well.  The calculations slightly
overestimate  the bandhead energy of  the $\gamma\gamma$-band; a similar result was also obtained for, e.g., $^{108}$Mo and $^{108}$Ru. The first q.p. $(h_{11/2})^2$ neutron alignment is predicted around $I=8$, and the transition to a 4-q.p. $nu(h_{11/2})^2 \pi (g_{9/2})^2$ band is expected to occur around $I=16$.

\begin{figure}[!htb]
\includegraphics[width=0.9\columnwidth]{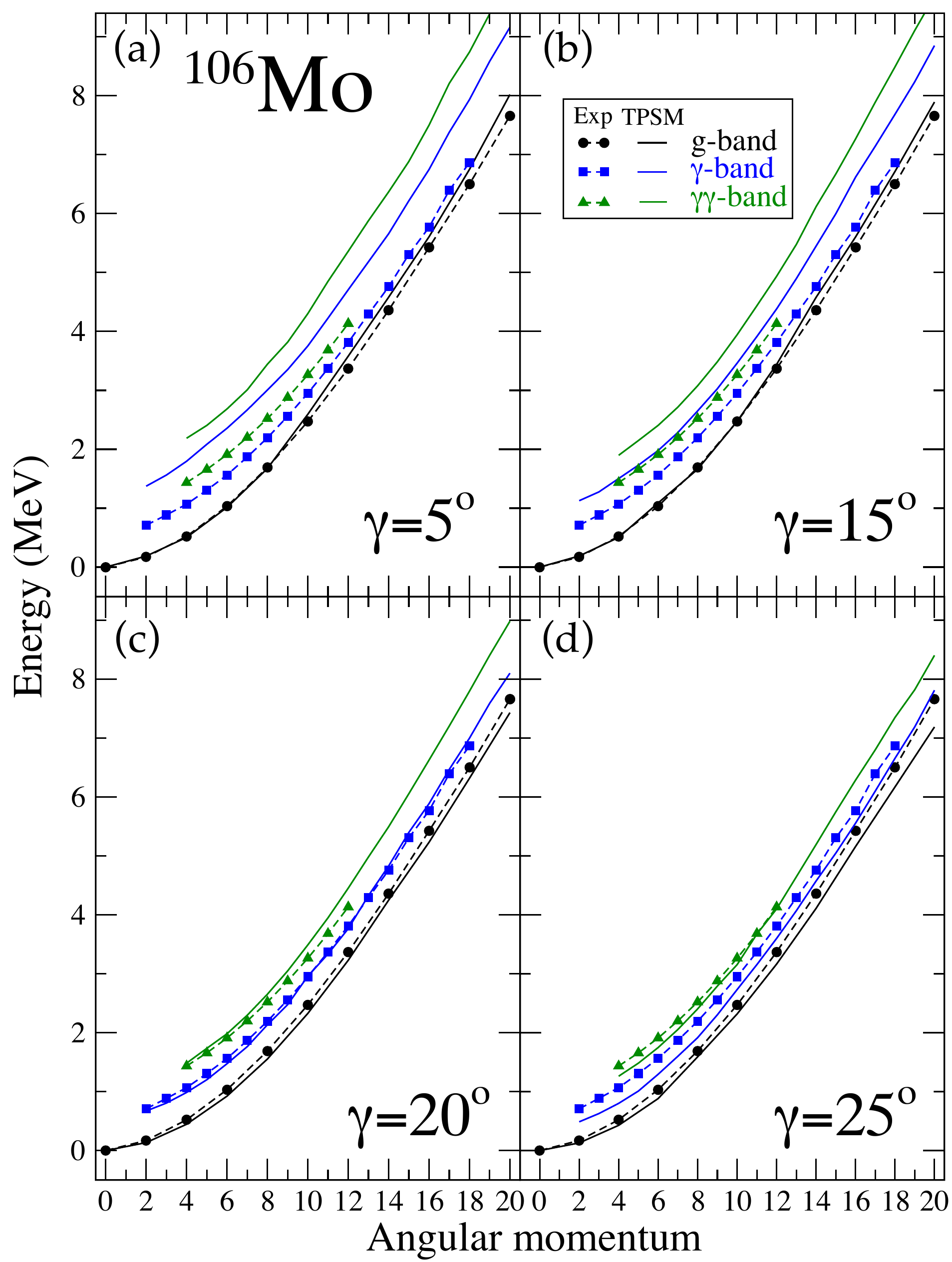}
\caption{(Color online) Comparison of experimental and calculated band
structures in $^{106}$Mo at four  values of triaxial deformation  $\gamma$.}
\label{fig12}
\end{figure}
%
To illustrate the importance of the $\gamma$ degree of freedom in the description
of the band structures in the Mo- and Ru-isotopes, we have carried out
TPSM calculations for  $^{106}$Mo for a range of $\gamma$-values. The obtained band structures at $\gamma = 5^\circ, 15^\circ, 20^\circ$, and $25^\circ$ are shown
in Fig.~\ref{fig12}. At $\gamma = 5^\circ$, 
 the calculated $\gamma$- and $\gamma\gamma$-bands are  shifted
with respect to experiment by more than 1\,MeV. At $\gamma = 15^\circ$,
$\gamma$- and $\gamma\gamma$-bands are shifted down in energy, and it is at $\gamma=20^\circ$ that all the three bands are reproduced fairly accurately. At still higher
value of $\gamma=25^\circ$, the deviation from experiment grows again. 
For $^{106}$Mo, therefore, $\gamma = 20^\circ$ is the optimum triaxial deformation in TPSM. Interestingly, the same value of $\gamma$ is predicted by CHFB+UNEDF0, see 
Table~\ref{tab0}.

\subsubsection{Transition Quadrupole Moments}

\begin{figure}[!htb]
\includegraphics[width=0.9\columnwidth]{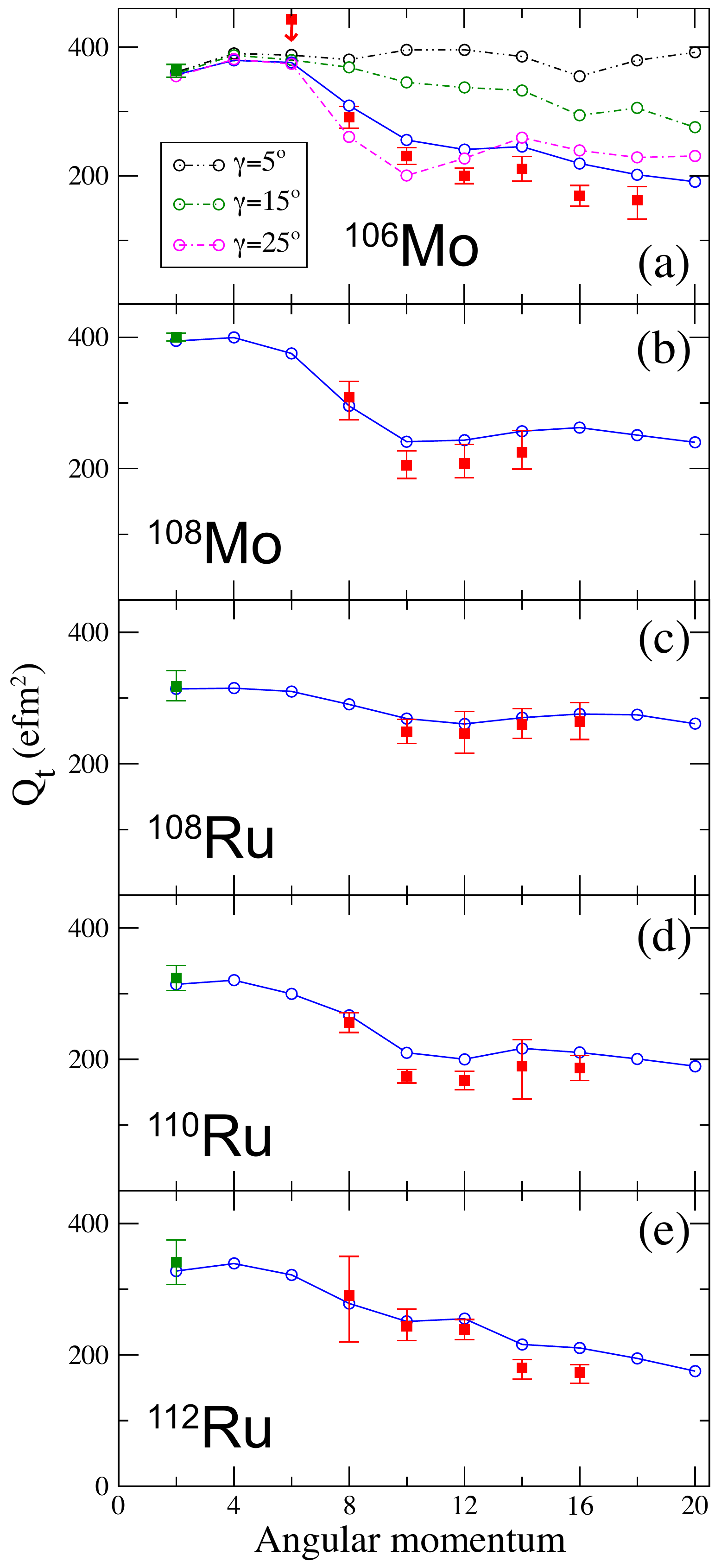}
\caption{(Color online) Transition quadrupole moments for $^{106,108}$Mo and  $^{ 108,110,112}$Ru
calculated in TPSM  (solid line) compared to experiment. The  $Q_t$ value  at $I=2$ is taken from  Ref.~\citep{Raman87} and  the high-spin values  from Ref.~\citep{Snyder13}.
For $^{106}$Mo, we also show TPSM results at $\gamma=5^\circ, 15^\circ$, and $25^\circ$ (dash-dotted lines).
}
\label{fig13}
\end{figure}
Using the TPSM  wave functions and  standard $E2$ effective charges 
($e_n=0.5e$ and $e_p=1.5e$)  we have evaluated the transition quadrupole moments  along the
yrast line of the studied Mo- and Ru-isotopes, see Fig.~\ref{fig13}. 
The overall behavior of the measured $Q_t$-values is reproduced quite well by the TPSM
approach. The drop in $Q_t$ observed for all the studied isotopes around $I=8$ is
due to the quasiparticle alignment of the $h_{11/2}$ neutron pair, and in some isotopes there
is a further drop around $I=16$ due to a consecutive alignment involving the $g_{9/2}$ proton pair. Figure~\ref{fig13}(a) also shows the TPSM predictions for $Q_t$ in $^{106}$Mo 
at $\gamma = 5^\circ, 15^\circ$, and $25^\circ$. Again, it is seen that the best reproduction of experimental data is obtained at the optimum value of $\gamma=20^\circ$.


\section{Conclusions}
\label{summary}

Stimulated by the recent experimental data on the transition quadrupole moments in the rotational bands of neutron-rich Mo and Ru nuclei \cite{Snyder13}, we studied the shapes of the  band structures in $^{106,108}$Mo, and $^{108,110,112}$Ru. We employed two complementary theoretical models: self-consistent CHFB+UNEDF0  approach  and TPSM. 

The triaxial PESs obtained in CHFB+UNEDF0 show stable g.s. triaxial minima in all cases. At higher angular momenta, the consecutive band crossings along the yrast line are expected, associated with the alignment of $\nu(h_{11/2})^2$ and $\pi(g_{9/2})^2$ pairs. The quadrupole deformations $\beta_2$ in the aligned bands are predicted to be reduced, but the shapes remain strongly triaxial. This result confirms predictions of an earlier works \cite{Skalski97,Faisal10} based on the cranked macroscopic-microscopic method with the Woods-Saxon average potential. The decrease of the corresponding transition quadrupole moments above  $\hbar\omega\sim 0.3$\,MeV reflects the change in  $\beta_2$ and $\gamma$  due to the q.p. alignment, and this reduction is consistent with experiment.

The results obtained with TPSM paint the same picture as CHFB calculations, and strongly favor the triaxial interpretation. Indeed, both the energies of yrast, $\gamma$, and $\gamma\gamma$ bands, and transition quadrupole moments are well described assuming stable triaxial shapes. Similar as in our CHFB calculations, transition quadrupole moments obtained in  TPSM exhibit a reduction    at neutron and proton band crossings.

In summary, according to our analysis, high-spin behavior of $^{106,108}$Mo, and $^{108,110,112}$Ru is consistent with  triaxial rotation. The predicted triaxial g.s.  minima are fairly shallow, and this perhaps is  why in some calculations, e.g.,  the cranked relativistic Hartree-Bogoliubov model of Ref.~\cite{Snyder13}, axial configurations may be slightly favored.

\begin{acknowledgments}
 
Useful discussions with Jacek Dobaczewski and Nicolas Schunck are gratefully acknowledged. This material is
based upon work supported by the U.S. Department of Energy, Office of
Science, Office of Nuclear Physics under Award Numbers No.\
DOE-DE-SC0013365 (Michigan State University) and No.\ DE-SC0008511
(NUCLEI SciDAC Collaboration). An award of
computer time was provided by the National Institute for
Computational Sciences (NICS) and the Innovative and Novel
Computational Impact on Theory and Experiment (INCITE) program using
resources of the OLCF facility.
\end{acknowledgments}


%

\end{document}